\documentclass{emulateapj}
\usepackage{apjfonts,amsmath}

\defcitealias{KK}{Paper I}
\newcommand{\paperI}{\citetalias{KK}}
\newcommand{\vct}[1]{\mathbf{#1}}
\newcommand{\bol}[1]{\mbox{\boldmath{$#1$}}}
\newcommand{\mach}{\mathcal{M}}
\newcommand{\step}{\mathcal{H}}
\newcommand{\Msun}{\rm M_\odot}
\newcommand{\kms}{\rm\;km\;s^{-1}}
\newcommand{\pcc}{\rm\;cm^{-3}}
\newcommand{\D}{\mathcal{D}}
\newcommand{\I}{\mathcal{I}}
\newcommand{\IRp}{\I_{1, R}}
\newcommand{\ITp}{\I_{1, \varphi}}
\newcommand{\IRc}{\I_{2, R}}
\newcommand{\ITc}{\I_{2, \varphi}}
\newcommand{\X}{f\,}
\newcommand{\rmin}{r_{\rm min}}
\newcommand{\ext}{{\rm ext}}
\newcommand{\cs}{c_{s}}

\shorttitle{DYNAMICAL FRICTION OF DOUBLE PERTURBERS}
\shortauthors{KIM, KIM, \& S{\'A}NCHEZ-SALCEDO}

\begin{document}
\title{Dynamical Friction of Double Perturbers in a Gaseous Medium}
\author{Hyosun Kim\altaffilmark{1}, Woong-Tae Kim\altaffilmark{1}, 
  and F.~J.~S{\'a}nchez-Salcedo\altaffilmark{2}}
\altaffiltext{1}{Department of Physics and Astronomy, FPRD, Seoul National 
  University, Seoul 151-742, Korea; hkim@astro.snu.ac.kr, wkim@astro.snu.ac.kr}
\altaffiltext{2}{Instituto de Astronom\'ia, Universidad Nacional Aut\'onoma 
de M\'exico, Ciudad Universitaria, 04510 Mexico City, Mexico; 
jsanchez@astroscu.unam.mx}

\begin{abstract}
In many astrophysical situations, as in the coalescence of supermassive 
black hole pairs at gas rich galactic nuclei,
the dynamical friction experienced by an object is a combination of its
own wake as well as the wakes of its companions.
Using a semi-analytic approach,
we investigate the composite wake due to, and the resulting drag forces on, 
double perturbers that are placed at the opposite sides of the 
orbital center and move on a circular orbit in a uniform gaseous medium.
The circular orbit makes the wake of each perturber asymmetric, creating
an overdense tail at the trailing side.  
The tail not only drags the perturber backward
but it also exerts a positive torque on 
the companion.  For equal-mass perturbers, the positive torque
created by the companion wake is, on average, a fraction $\sim40-50$\% of 
the negative torque created by its own wake, but this fraction may be even 
larger for perturbers moving subsonically.  This suggests that
the orbital decay of a perturber in a double system, especially in the 
subsonic regime, can take considerably longer than in isolation.
We provide the fitting formulae for the
forces due to the companion wake and discuss our results in light of 
recent numerical simulations for mergers of binary black holes.
\end{abstract}


%------------------------------------------------------------------------------
\section{INTRODUCTION}

Understanding the nature of the dynamical friction (DF) force 
is of great importance to describe the evolution of
gravitational systems. 
Although the concept of DF was introduced by \citet{chandra} 
studying collisionless backgrounds, it is also of astrophysical interest 
when considering gaseous media 
(e.g, \citealt{dok64,rud71,rep80,ost99,san99,san01}).  
In a seminal paper, \citet{ost99} derived the analytic formulae for 
drag forces on a perturber moving straight 
in a uniform gaseous medium, which have found a variety
of astrophysical applications, 
from accretion disks (e.g. \citealt{nar00,kar01,cha01}) to
the intracluster medium (e.g., \citealt{kim05,kim07,con08}).
Recently, \citet[hereafter \paperI]{KK} extended the work of
\citet{ost99} to a more realistic case where the 
perturber moves on a circular orbit (see \citealt{bar07} for
the relativistic case).  

Although the works mentioned above have improved our understanding on
DF force on a single object, 
there are many astronomical situations involving double or multiple bodies 
in which one needs an analytical estimate of the net drag force.
While the orbital evolution of a binary system due to the DF in
a collisionless system has been studied in great detail 
(e.g., \citealt{heg75}), it is still lacking for the gaseous case.
The latter is crucial to describe the formation and hardening
of close binary systems \citep{bat02},
merging of double black holes at galactic centers 
\citep{esc04,esc05,dot06,dot07,may07}, and 
orbital decays of kpc-sized giant clumps formed 
in primordial ``clump cluster'' galaxies \citep{imm04,bou07}. 
In particular, angular momentum loss to gas provides a plausible mechanism
to explain the relatively rapid coalescence of supermassive 
black hole pairs in galaxy centers (e.g., \citealt{beg80,gou00,arm02,arm05}).
\citet{esc04,esc05} found, through numerical simulations, that 
the black holes produce a composite wake of an inclined ellipsoidal
shape that exerts a net torque on the black holes. 
This clearly indicates that the in-spiral of one black hole is 
affected also by the wake from its companion.
In this \textit{Letter}, we consider a system composed
of two perturbers on coplanar circular orbits 
in order to assess quantitatively the effect of the companion wake.
Using a semi-analytic 
approach, we study the structure of the combined density wake, 
evaluate the resulting drag force on each object, 
and apply our results to the cases considered 
in numerical simulations of DF-induced mergers of black holes.

%---------------------------------------------------------------------------
\section{FORMULATION}

We consider two point-mass perturbers moving on coplanar
circular orbits around
the same orbital center in an inviscid gaseous medium.\footnote{The
perturbers may orbit under a common external gravitational potential
and/or under their mutual gravity.
They constitute a binary if the latter dominates.}
In order to isolate the effects of the companion, we assume a perfectly
uniform background with density $\rho_0$ and ignore the
complications from density gradients.
To simplify the presentation further, we consider that both perturbers
orbit with a fixed radius $R_p$ and at a constant velocity $V_p$,
and are located at the opposite sides of the orbital center,
but the extension to general case is straightforward.
To study the response of gas to the perturbers, we employ the same
formalism as in Paper I for a time-dependent
linear perturbation analysis.
The reader is referred to Paper I for a more detailed description
of the method.

Assuming that the perturbed density $\alpha = (\rho-\rho_0)/\rho_0$
is adiabatic and very small, we linearize the equations of hydrodynamics
to obtain a three-dimensional wave equation
\begin{equation}\label{eq:wave}
  \vct{\nabla}^2\alpha-\frac{1}{\cs^2}\frac{\partial^2\alpha}{\partial t^2}
  = -\frac{4\pi G}{\cs^2}\rho_\ext(\vct{x},t),
\end{equation}
where $\cs$ is the adiabatic speed of sound in the unperturbed medium
and $\rho_\ext$ denotes the mass density of the perturbers 
(e.g., \citealt{ost99}). 

We work in cylindrical coordinates $(R,\varphi, z)$ with its origin lying 
at the orbital center and the $z$-axis perpendicular to the orbital plane.
Assuming that the perturbers with mass $M_p$ and $fM_p$ each
(with $f$ denoting the mass ratio) are introduced at
$(R_p, 0, 0)$ and $(R_p, \pi, 0)$ at $t=0$, respectively, one can write
\begin{eqnarray}\label{eq:rho_ext}
  \rho_\ext(\vct{x},t)&=& M_p\,\step (t)\,\delta (R-R_p)\,\delta (z)\nonumber\\
  &\times& 
  \{\delta[R_p\,(\varphi-\Omega t)]+\X \delta[R_p\,(\varphi-\pi-\Omega t)]\},
\end{eqnarray}
where $\step (t)$ is the Heaviside step function
and $\Omega \equiv V_p/R_p$ is the angular speed of the perturbers.  
Since equation (\ref{eq:wave}) is linear, $\alpha$ is given 
by a simple superposition of the wakes of both perturbers.
By solving equation (\ref{eq:wave}) based on the retarded Green's function 
technique and simplifying the resulting integral analytically,
one can show that the perturbed density is reduced to
\begin{equation}\label{eq:alpha}
  \alpha (\vct{x},t) = \frac{G M_p}{\cs^2 R_p}\, \D (R,\varphi,z,t),
\end{equation}
where $\D=\D_1+ \X \D_2$ with 
$\D_1$ and $\D_2$ denoting the dimensionless wake, 
given by equation (8) in Paper I, of the perturber with 
mass $M_p$ and $fM_p$, respectively.  
Note that $\D_2(R,\varphi,z,t)=\D_1(R,\varphi-\pi,z,t)$
for the perturbers in consideration.

The gravitational drag force exerted on the perturber of mass $M_p$
located at the position $\vct{x}_p$ 
can be obtained by directly evaluating the integral
\begin{equation}\label{eq:DF}
  \vct{F}_\mathrm{DF} = G M_p \rho_0 \int d^3\vct{x}\ 
  \frac{\alpha(\vct{x},t)\ (\vct{x}-\vct{x}_p)} 
       {\vert \vct{x}-\vct{x}_p \vert^3} = 
  \vct{F}_{\rm DF,1} + 
  \vct{F}_{\rm DF,2},
\end{equation}
where $\vct{F}_{\rm DF,1} = -\mathcal{F}
  (\IRp\,\hat{\vct{R}} + \ITp\,\hat{\bol{\varphi}})$ 
and $\vct{F}_{\rm DF,2} = -\mathcal{F}\X
  (\IRc\,\hat{\vct{R}} + \ITc\,\hat{\bol{\varphi}})$,
with $\mathcal{F} \equiv 4\pi\rho_0\,(GM_p/V_p)^2$.
Here, $\hat{\vct{R}}$ represents the
unitary direction vector along $R$, and
$\IRp$ and $\ITp$ are the dimensionless 
drag forces on a perturber by its own wake 
in the radial and azimuthal directions, respectively, 
while $\IRc$ and $\ITc$ refer to those
from the wake of its companion with mass $fM_p$. 
The former was evaluated and widely discussed in \paperI. 
We here focus on $\IRc$ and $\ITc$ defined by equations (12) of
\paperI\ except $\D_2$ replacing $\D$ for the perturbed density.

As in Paper I, we calculate $\D(\vct{x},t)$ and
$\vct{F}_\mathrm{DF}$ on a three-dimensional Cartesian mesh centered
at the center of the orbit.
We checked that the grid spacing of $\sim R_p/640$ and the box
size of $\sim$(20--100) $R_p$ are sufficient to give converged
results.

%------------------------------------------------------------------------------
\section{RESULTS}

%------------------------------------------------------------------------------
\subsection{Density Wake}\label{sec:wake}

%fig1 : 14
\begin{figure}%fig1
  \epsscale{1.2}
  \plotone{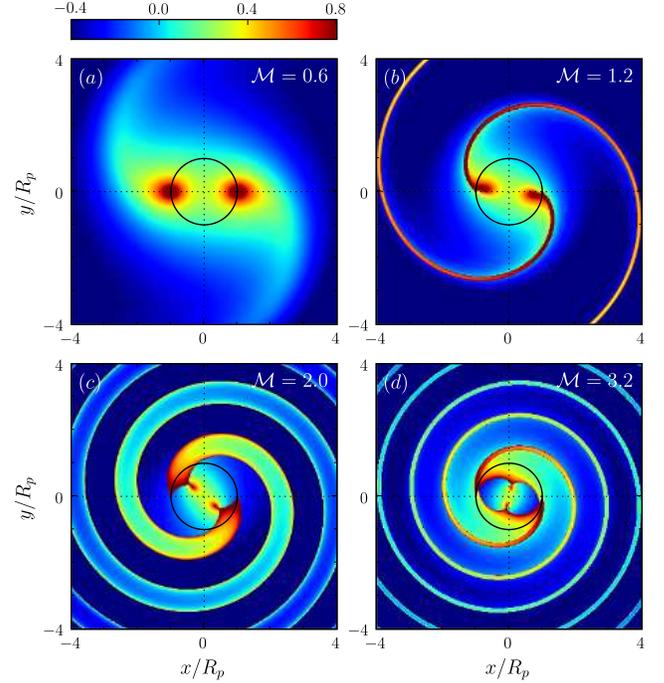}
  \caption{\label{fig:wake}
  Distributions of the dimensionless perturbed density $\D$
  in a steady state for 
  (\textit{a}) $\mach=0.6$,
  (\textit{b}) 1.2, (\textit{c}) 2.0, and (\textit{d}) 3.2
  on the orbital plane ($z=0$). 
  The perturbers of equal mass located at $(x,\,y)=(\pm R_p,\,0)$
  are moving in the counterclockwise direction 
  along a circular orbit marked by the black circle in each frame.
  Colorbar labels $\log \D$.}
\end{figure}

In this section, we limit our presentation to the cases with equal-mass 
perturbers; the cases with $f\neq 1$ will be 
briefly discussed in \S\ref{sec:dis}.
As the perturbers introduced at $t=0$ move along a circular orbit,
they continuously launch sound waves that propagate and affect the
surrounding medium that would otherwise be uniform and static.  
Any location inside the causal region is able to
receive sonic perturbations from both perturbers, possibly multiple times, 
creating a density wake that differs significantly depending on the 
Mach number $\mach\equiv V_p/\cs$ of the perturbers. 
Figure \ref{fig:wake} displays the distributions of
the dimensionless wake $\D$ on the orbital plane ($z=0$),
when a steady state is reached ($t \gg R_p/\cs$) 
for $\mach=0.6$, 1.2, 2.0, and 3.2.  
In each frame, the perturbers located at $(x,y)=(\pm R_p,0)$ are 
rotating counterclockwise. 

For subsonic perturbers ($\mach<1$), the perturbed density is smooth
without involving a shock.  The bending of the wakes caused by the circular
motions leads to slight over-densities at the trailing sides
(see Fig.~\ref{fig:wake}a), producing nonvanishing drag forces.
In the steady state, which is achieved at $t\rightarrow\infty$,
a parcel of gas at any position receives one
sonic perturbation from each perturber.
For supersonic cases ($\mach>1$), on the other hand, 
the wake of each perturber initially consists of a sonic sphere and 
a Mach cone, the interiors of which are influenced 
by sonic disturbances once and twice, respectively.
Because of the circular motion, a perturber (and the head of its Mach cone)
is able to overtake its own sonic sphere and subsequently the other 
sonic sphere from the companion, both of which are expanding radially outward.
This in turn provides additional perturbations to the wakes and thus forms
long high-density tails that loosely wrap the 
perturbers in a trailing spiral fashion, as Figure~\ref{fig:wake} shows.  
The tails in fact trace the regions bounded by shock discontinuities
where the gas has received sonic signals four times 
(three from one perturber and one from the other) and 
do not overlap with each other,
provided the Mach number is less than 2.972 (see below).  
Note that the densest parts of
a tail are located at the immediate trailing side of a perturber, 
which indicates that the companion wake generally tends to reduce 
the net DF force, as we shall show in \S\ref{sec:force}.

The wake tails thicken as the Mach number increases from unity.
\paperI\ showed that a tail from a single perturber becomes fat 
enough to make the inner edge contact with the outer edge at 
$\mach_1=4.603$.  The self-overlapping of a tail develops 
a new thin tail that becomes thicker with
increasing $\mach$ and again overlaps itself at $\mach_2=7.790$, and so on.  
In double-perturber cases, however, one tail is able to 
mutually overlap with the other even before it undergoes 
self-overlapping.
The critical Mach numbers $\mach_n$ for the mutual 
overlapping of tails are determined by equation (B3) in \paperI\ 
for half-integer $n$.  A few critical Mach numbers are
$\mach_{1/2}=2.972$, $\mach_{3/2}=6.202$, and $\mach_{5/2}=9.371$.
The high-density narrow tails shown in Figure~\ref{fig:wake}d for 
$\mach=3.2$ are constructed by combining six sonic disturbances
emitted by the perturbers.

%------------------------------------------------------------------------------
\subsection{Gravitational Drag Force}\label{sec:force}

\begin{figure}%fig2
  \epsscale{1.04}
  \plotone{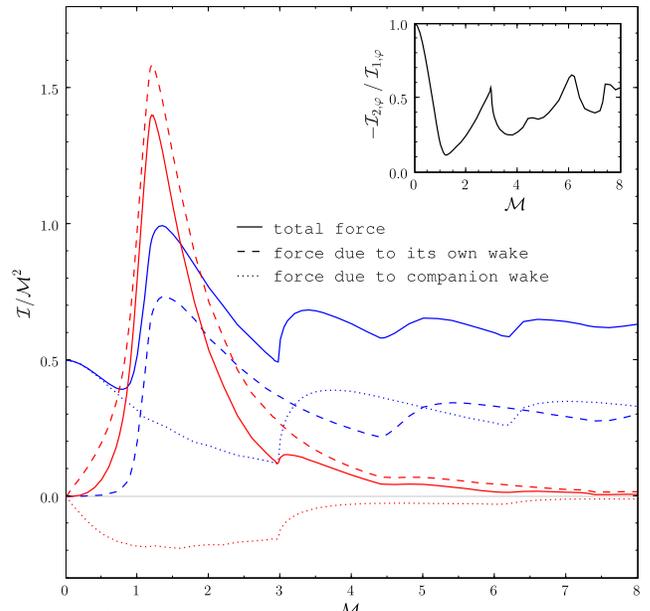}
  \caption{\label{fig:force}
  Gravitational drag forces on a perturber in a double system 
  in the radial (\textit{blue}) and azimuthal (\textit{red}) directions 
  as functions of the Mach number $\mach$. 
  The dashed curves adopted from Paper I with $\rmin/R_p=0.1$ 
  represent the forces ($\I_1$) 
  originated from the wake of the perturber itself,
  while those ($\I_2$) from the companion wake are plotted as dotted lines.
  The solid lines give the net DF forces ($\I_1+\I_2$) for
  equal-mass perturbers.  The inset plots the ratio $-\ITc/\ITp$ 
  which is positive and less than unity for all $\mach$.}
\end{figure}

Once the perturbed density $\D(\vct{x},t)$ is constructed, 
it is straightforward to calculate the DF forces exerted on each perturber.  
While the volume of
space influenced by the perturbers steadily increases with time,
the resulting drag forces quickly converge to their steady-state
values typically within one orbital period.  
Figure~\ref{fig:force} plots the various drag forces in a steady state
on a perturber as functions of the Mach number.
In order to avoid a divergence of the force integral, 
only the region with the distance $r>\rmin=R_p/10$ from the perturber 
is taken into account in the force computation. 
Note that only $\ITp$ depends on the Coulomb logarithm $\ln (\rmin/R_p)$ for 
supersonic perturbers; the other three forces
($\IRp$, $\IRc$, and $\ITc$) are independent of the adopted value for 
$\rmin$.  
The local bumps in $\IRc$ and $\ITc$ are caused by 
the overlapping of the tails occurring at the critical Mach numbers,
as discussed in \S\ref{sec:wake}.

Figure~\ref{fig:force} shows that $\ITc$ has opposite sign to 
$\ITp$ for all $\mach$.  This implies that, regardless of the Mach number,
one perturber in a double-perturber system gains 
angular momentum from the gravitational torque exerted by the companion 
wake, while its own wake always takes away angular momentum from it.  
For equal-mass perturbers,
the net drag force in the azimuthal direction is thus smaller than the 
isolated counterpart.
The contribution of the companion wake to the DF force in the azimuthal 
direction is delineated in the inset of Figure~\ref{fig:force}. 
In the supersonic range, the ratio $-\ITc/\ITp$ is on average $\sim40\%$, 
and drops to $\sim12\%$ at $\mach\sim1.2-1.4$
where the net azimuthal drag force is maximized. 
Since $\ITp$ increases with decreasing $\ln(\rmin/R_p)$ for $\mach>1$, 
the effect of the companion wake on the orbital
decay of supersonic double perturbers would diminish
as the perturber size nominally represented by $\rmin$ 
decreases relative to the orbital radius.
Interestingly, in the subsonic case, the ratio $-\ITc/\ITp$,
which does not depend on $\rmin$, 
is $\sim 50\%$ at $\mach \sim 0.7$ and steeply 
approaches unity as $\mach\rightarrow 0$,
suggesting that the effect of the companion
wake is larger as the speed of perturbers decreases.

%fig2 : 14

On the other hand, $\IRp$ and $\IRc$ are of comparable amplitude
over a wide range of $\mach$, and thus give rise to a net radial drag 
on double perturbers that is about twice larger than in the 
corresponding single-perturber 
cases\footnote{When $\mach=0$,
the steady-state solution of equation (\ref{eq:wave}) is simply 
$\alpha=(GM_p/\cs^2) (|r-R_p|^{-1}+f|r+R_p|^{-1})$, for which 
equation (\ref{eq:DF}) yields
$\IRc/\mach^2 = 0.5$ and $\ITc/\mach^2 =0$.}. 
They affect the orbital eccentricity rather than removing angular momentum
much (e.g., Paper I), which
may have a gravitational wave signature detectable with
\textit{Laser Interferometer Space Antenna (LISA)} \citep{arm05}.

For practical purposes, we fit our results for $\IRc$ and $\ITc$ using
\begin{equation}\label{eq:IRc}
  \frac{\IRc}{\mach^2} = 
  \left\{\begin{array}{l@{\ \textrm{if}\ }l}
  0.5-0.43\Big(1-\cosh^{-0.36} (2.2\mach)\Big) & \mach\!\!<\! 2.97,\\
  0.76-0.08\Big(\mach+(\mach-2.76)^{-1}\Big) & 2.97\!\!\leq\!\mach\!\!<\!6.2,\\
  0.56-0.027\Big(\mach+(\mach-6)^{-1}\Big) & \mach\!\!\geq\! 6.2,
  \end{array}\right.
\end{equation}
and
\begin{equation}\label{eq:ITc}
  \frac{\ITc}{\mach^2} =
  \left\{\begin{array}{l@{\quad\textrm{if}\ \mach}l@{\,2.97,}}
  -0.022\ (10-\mach)\;\tanh\,(3\mach/2)  &<\\
  -0.13+0.07\,\tan^{-1} (5\mach-15)      &\geq
  \end{array}\right.
\end{equation}
which are accurate within 6\% of the numerical results for all $\mach$.
The algebraic fits to $\IRp$ and $\ITp$ are given by equations (13) and (14) 
of \paperI.  It can be shown that
$\ITp\approx-\ITc\rightarrow \mach^3/3$
in the limit of small $\mach$.

%------------------------------------------------------------------------------
\section{DISCUSSION}\label{sec:dis}

For a system composed of two perturbers moving on coplanar circular orbits
at the opposite sides of the system center, 
we have found that a perturber is dragged backward by its own 
induced wake, while it is simultaneously pulled forward by the 
wake of its companion.  
For equal-mass perturbers, 
the ratio of the positive torque from the companion wake to the
negative torque from its own wake varies between 0.1 to 1.0,
and has a mean value at about 0.4.
This indicates that, since the wake tails are a large-scale
perturbation, the effect of a companion wake on the 
orbital decay of a double system is by no means negligible
except perhaps at $\mach\sim 1.2$--$1.4$.  
When the perturbers do not have the same orbital radius, 
the positive torque generated by the companion wake 
is expected to be enhanced (thus making the net azimuthal drag reduced)
if the companion has a larger orbital radius.
In order to quantify this effect, we have computed the drag force, 
for instance, on a body with $\mach=1$ at $R_p$ when the companion
is at $2R_p$ and $\mach=2$ (in order to have the same $\Omega$), 
and found that the positive torque increases by a factor of $\sim1.4$.
If the perturbers have different orbital frequencies so that one perturber
completes several orbits in the orbital period of the companion, 
the orbit-averaged torque exerted by the wake of the companion is likely to 
be reduced.
In the limit of very different frequencies, the orbit-averaged
torque by the companion wake becomes negligible.

The results of this \textit{Letter} can be immediately applied to the 
numerical models considered in \citet{esc04} for DF-induced mergers 
of supermassive black holes in a gaseous medium.
In their model, two black holes of equal mass are initially 
separated from each other widely and orbit at near-transonic speed 
under an external gravitational potential and begin to undergo 
gaseous drag.
Our results shown in Figure~\ref{fig:force} suggest that 
the companion wake presumably plays a minor role in this early phase 
of the orbital decay.
\citet{esc04} reported that at some point when the binary separation 
is reduced to $\sim7$ pc, the binary produces a wake in the surrounding 
medium that is well approximated by an ellipsoid with an axis ratio of 2:1.  
The major axis of the ellipsoid lags behind the binary axis 
by $\sim22.5\arcdeg$.  Since the wake configuration of this sort
can be obtained by blurring the perturbed density shown in 
Figure~\ref{fig:wake}a, we infer that the black holes have
$\mach\sim 0.6$ at this time.
The binary keeps hardening by an ellipsoidal torque, and 
the effect of the companion wake is now non-negligible at all. 

%fig3 : 13
\begin{figure}%fig3
  \epsscale{1.1}
  \plotone{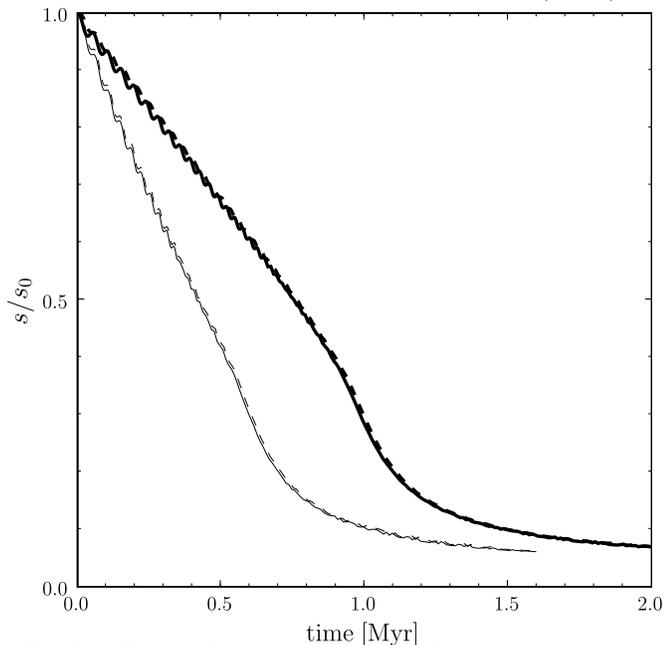}
  \caption{\label{fig:ex}
Decay of the separation of a black hole binary of equal mass caused
by dynamical friction due to a background gas.  
The thick and thin
curves correspond to the cases when the wakes of
both  binary components are considered and when the companion wake
is ignored, respectively.  In both cases, the solid lines plot the 
results based on both radial and azimuthal forces, while those with
only azimuthal forces are given as dashed lines.
}
\end{figure}

Figure 10 of \citet{esc04} shows that it takes the binary about 
1.5 Myr to decay from 7 to 0.7 pc.  To check whether this is 
consistent with our predictions,  we consider a
binary black hole with mass $M_p=5\times 10^8\Msun$ each,
embedded in a uniform medium with number density $n_0=1.5\times 10^5\pcc$
and sound speed $\cs=650\kms$, a condition similar to those in \citet{esc04}.
Initially, the binary has $\mach=0.6$ and a separation of $s_0=7$ pc.
We follow the orbital decay of the binary subject to the DF forces
found in \S\ref{sec:force}.
Figure~\ref{fig:ex} plots the resulting temporal changes of the
separation $s$.
The thick curves correspond to the cases
when the wakes of both binary components are considered, which show that
the binary decays to $s=0.7$ pc in $\sim1.5$ Myr, consistent with the
results of \citet{esc04}.
When the effects of the companion wake are neglected, 
the decay time becomes shorter by about a factor of 1.5, 
as indicated by the thin lines.
Note that the close agreement between the two cases with and without
the radial force demonstrates that it does not affect the decay much.

The fact that the companion wake always acts against the orbital decay of
a double system suggests an interesting possibility 
that a less massive perturber in an unequal-mass system 
may be able to experience a forward net force and move radially outward 
by acquiring (instead of losing) angular momentum from the wake of
a more massive body, resembling a dynamical barrier,
while the latter is little influenced by the former.  
This \textit{dynamical boost} happens if the mass ratio $f$ of two 
components exceeds $|\ITp/\ITc|$.
The evolution of unequal-mass perturbers is more complex
since the assumption of identical orbits for both perturbers
will fail soon.  In practice, one expects that the orbit of the light 
perturber will be quenched to the massive perturber;  
the former is trapped on a radius where the drag of
its own wake is slightly larger than the forward force of the
companion wake, until the orbit of the massive perturber decays
and the wakes decouple.  We will discuss this case somewhere else.

%------------------------------------------------------------------------------
\acknowledgments

We are grateful to an anonymous referee for a thoughtful report.
This work was supported by KICOS through the grant
K20702020016-07E0200-01610 provided by MOST,
and partly for H.K.~by the BK21 project of the Korean Government.
The computations were performed on the
Linux cluster at the KASI built with funding from KASI and 
the ARCSEC.
F.J.S.S.~acknowledges financial support from projects 
PAPIIT IN114107 and CONACyT 60526.

\clearpage

\end{document}